\documentclass[aps,prl,twocolumn,showpacs,floatfix,preprintnumbers,nofootinbib,superscriptaddress]{revtex4}

\usepackage{graphicx}
\usepackage{color}
\usepackage{natbib}
\usepackage{multirow}

\begin{document}

\title{Cosmology seeking friendship with sterile neutrinos}

\author{Jan Hamann}
\affiliation{Department of Physics and Astronomy,
 University of Aarhus, 8000 Aarhus C, Denmark}

\author{Steen Hannestad}
\affiliation{Department of Physics and Astronomy,
 University of Aarhus, 8000 Aarhus C, Denmark}

\author{Georg~G.~Raffelt}
\affiliation{Max-Planck-Institut f\"ur Physik
  (Werner-Heisenberg-Institut),
  F\"ohringer Ring~6, 80805 M\"unchen, Germany}

\author{Irene Tamborra}
\affiliation{Max-Planck-Institut f\"ur Physik
 (Werner-Heisenberg-Institut),
 F\"ohringer Ring~6, 80805 M\"unchen, Germany}
\affiliation{Dipartimento Interateneo di Fisica
``Michelangelo Merlin'', Via Amendola 173, 70126 Bari, Italy}
\affiliation{INFN, Sezione di Bari, Via Orabona 4, 70126 Bari, Italy}

\author{Yvonne Y.~Y.~Wong} \affiliation{Institut f\"ur Theoretische
  Teilchenphysik und Kosmologie, RWTH Aachen, 52056 Aachen, Germany}

\date{25 June 2010, revised 23 September 2010}

\preprint{MPP-2010-73, TTK-10-36}

\begin{abstract}
  Precision cosmology and big-bang nucleosynthesis mildly favor extra
  radiation in the universe beyond photons and ordinary neutrinos,
  lending support to the existence of low-mass sterile neutrinos.  We
  use the WMAP 7-year data, small-scale CMB observations from
  ACBAR, BICEP and QuAD, the SDSS 7th data release,
  and measurement of the Hubble parameter from HST
  observations to derive credible regions for the assumed
  common mass scale $m_{\rm s}$ and effective number $N_{\rm s}$ of
  thermally excited sterile neutrino states.
  Our results are compatible with the existence of one or perhaps two
  sterile neutrinos, as suggested by LSND and MiniBooNE,
  if $m_{\rm s}$ is in the sub-eV range.
\end{abstract}


\maketitle

{\em Introduction.}---Big-bang nucleosynthesis (BBN) has for many
years provided the best inference of the radiation content of the
universe. This inference implies the existence of approximately
three neutrino flavors and hence three generations of elementary
particles~\cite{Dolgov:2002wy,Iocco:2008va}.  Of course, by the
early 1990s, precision measurements of the $Z^0$ decay width at LEP
had superseded this early cosmological evidence.  More recently,
observations of the cosmic microwave background (CMB) anisotropies
and the large-scale structure (LSS) distribution have allowed us to
probe the radiation density at the CMB decoupling epoch.  Since this
epoch corresponds to a much later time than the epoch of
nucleosynthesis, inference of the radiation content of the universe
from precision CMB/LSS measurements provides an important overall
consistency check. The results are surprising on several counts.

Firstly, when analyzing cosmological data in extended scenarios that
allow for extra radiation, one consistently finds a trend towards a
radiation excess, and this hint has become stronger over the
years~\cite{Hamann:2007pi,Hamann:2010pw,GonzalezGarcia:2010un}. The
cosmic radiation content is usually expressed in terms of $N_{\rm
  eff}$, representing the effective number of thermally excited
neutrino species. The standard value $N_{\rm eff}=3.046$
slightly exceeds 3 because $e^+e^-$ annihilation provides residual
neutrino heating~\cite{Mangano:2005cc}. Most recently, the Wilkinson
Microwave Anisotropy Probe (WMAP) collaboration found a 68\%-credible
interval of $N_{\rm eff} = 4.34^{+0.86}_{-0.88}$ based on their 7-year
data release and additional LSS data~\cite{Komatsu:2010fb}.  A similar
study including the Sloan Digital Sky Survey (SDSS) data release 7
(DR7) halo power spectrum found $N_{\rm eff} = 4.78^{+1.86}_{-1.79}$
at 95\% confidence \cite{Hamann:2010pw}; see also
Ref.~\cite{Reid:2009nq}.  Most recently the Atacama Cosmology Telecope
Collaboration has arrived at similar
conclusions~\cite{Dunkley:2010ge}.

Secondly, for a good part of the past two decades, BBN and the
observed primordial $^4$He abundance suggested $N_{\rm eff} \lesssim
3$ \cite{Simha:2008zj}. A large body of literature was devoted to
particle-physics constraints, assuming that additional radiation
required other novel ingredients such as a $\nu_e$--$\bar\nu_e$
asymmetry~\cite{Dolgov:2002wy,Iocco:2008va}.  However, the BBN
situation has also changed of late: Two recent studies find a somewhat
higher $^4$He abundance of $Y_{\rm p} = 0.2565 \pm 0.001 \, ({\rm
  stat}) \pm 0.005 \, ({\rm syst})$ \cite{Izotov:2010ca} and $Y_{\rm
  p} = 0.2561 \pm 0.011$ \cite{Aver:2010wq} respectively, suggesting
additional radiation during the BBN epoch.

Taking these hints seriously, one quickly finds that it is not easy
to account for additional radiation because
well-motivated candidates are lacking. The strong dilution of any pre-existing
radiation at the quark-hadron phase transition at $T\sim 170$~MeV
suggests that a novel radiation component should be produced after
the QCD epoch, although there are
counter-examples~\cite{Chun:2000jr}. If neutrinos are Dirac
particles there is no obvious way to equilibrate the right-handed
states because one needs larger couplings (e.g.,
magnetic moments or right-handed currents) than allowed by other
evidence~\cite{Raffelt:1999tx}. Thermal axions can provide only a
fractional contribution to $N_{\rm eff}$ \cite{Hannestad:2005df}. In
contrast to earlier findings, it was recently shown that standard
neutrinos with primordial asymmetries can provide any value for
$N_{\rm eff}$ \cite{Pastor:2008ti}. No simple mechanism,
however, produces the required asymmetries.

If low-mass sterile neutrinos exist and mix with active flavors, they
can be thermally excited by the interplay of oscillations and
collisions~\cite{Kainulainen:1990ds}. The prior theoretical
motivation for such states is perhaps not strong. However, the signal
for $\bar\nu_\mu\to\bar\nu_e$ flavor conversion in the LSND
experiment~\cite{Aguilar:2001ty}, if interpreted in terms of flavor
oscillations, requires low-mass sterile states
\cite{GonzalezGarcia:2007ib,Strumia:2002fw,Cirelli:2004cz}. Later,
the MiniBooNE experiment observed excess events in the
$\nu_\mu\to\nu_e$ channel that were not compatible with LSND in a
two-flavor oscillation model~\cite{AguilarArevalo:2007it,
AguilarArevalo:2008rc}. Early MiniBooNE data in the
$\bar\nu_\mu\to\bar\nu_e$ channel were statistically not significant
enough to confirm or refute the LSND
signature~\cite{AguilarArevalo:2009xn}. However, based on 70\% more
data, a clear excess was recently reported~\cite{MB2010}. If
interpreted in terms of flavor oscillations, these data require CP
violation and thus a minimal scenario would involve 2 sterile
neutrinos that can interfere
appropriately~\cite{Karagiorgi:2009nb,Sterile2010},
although this interpretation causes significant tension with
disappearance experiments, notably atmospheric neutrinos. An
alternative is 1 sterile state together with nonstandard
interactions~\cite{Akhmedov:2010vy}. Yet another motivation for
low-mass sterile neutrinos arises from a combined fit of MiniBooNE
and Gallium radioactive source experiments~\cite{Giunti:2010wz}.

These scenarios together with the cosmological hints for
extra radiation motivate us to reconsider the impact of low-mass
sterile neutrinos on cosmological  observables.  Some recent
studies have looked at light sterile states in the context of
precision cosmology~\cite{Dodelson:2005tp, Melchiorri:2008gq,
Acero:2008rh}, but with emphasis on providing constraints on such
models. Furthermore, these papers were written at a time when
precision cosmology provided only very weak evidence for $N_{\rm eff}
> 3$.

{\it Scenarios.}---Thermal excitation of sterile neutrinos is
complicated because of possible resonant matter effects and CP
violating phases. However, for the eV mass range and the relatively
large mixing angles needed to explain the oscillation experiments,
the sterile states are likely strongly thermally
excited~\cite{Melchiorri:2008gq,DiBari:2001ua}.
Note however that  the presence of even a small neutrino asymmetry can
lead to incomplete  thermalization~\cite{Foot:1995bm,fuller}.

In the following we take the number of thermally excited
sterile neutrinos, $N_{\rm s}$, to be an adjustable cosmological fit
parameter, while the active neutrinos are assumed to have a fixed
abundance ($N_\nu=3.046$) so that $N_{\rm
  eff}=3.046+N_{\rm s}$.  For the neutrino masses, we consider two
schematic scenarios.  In the first scenario, the ordinary neutrinos
are taken to be
massless, while the sterile states have a common mass
scale $m_{\rm s}$ which is free to vary  (the ``$3+N_{\rm
  s}$'' scenario).  The second scenario, the ``$N_{\rm s}+3$'' scenario,
consists of massless sterile states and active
neutrinos that have an adjustable common mass $m_\nu$, i.e., all
active species are treated as degenerate in mass.  This is a good
approximation because current cosmological data are not sensitive to
the small mass splittings.

{\it Cosmological analysis.}---Besides the new parameters $N_{\rm
s}$ and $m_{\rm s}$ or $m_{\nu}$, we use a cosmological parameter
space consisting of the standard ``vanilla'' $\Lambda$CDM
parameters: The baryon density $\omega_{\rm b}=\Omega_{\rm b} h^2$,
cold dark matter density $\omega_{\rm cdm}=\Omega_{\rm cdm} h^2$,
Hubble parameter $H_0$, scalar fluctuation amplitude $A_{\rm s}$,
scalar spectral index $n_{\rm s}$, and the optical depth to
reionization $\tau$.  We use either $m_\nu$ or $m_{\rm s}$ as a fit
parameter, depending on the scenario under consideration,
from which we calculate the contribution to the matter
density as (i) $\Omega_\nu h^2 = N_{\rm s} \times m_{\rm s}/(93 \ {\rm
eV})$ in the $3+N_{\rm s}$ scheme, and (ii) $\Omega_\nu h^2 = 3.046
\times m_\nu/(93 \ {\rm eV})$ in the $N_{\rm
  s}+3$ case.  We impose flat priors on all parameters, as detailed in
Table~\ref{tab:priors}.  Parameter estimation is performed using a
modified version of the \texttt{CosmoMC} package~\cite{Lewis:2002ah}.
\begin{table}
\vspace*{-0.2cm}
\caption{Priors and standard values for the cosmological fit
parameters considered in this work. \label{tab:priors}}
\begin{ruledtabular}
\begin{tabular}{lll}
Parameter& Standard & Prior \\
\hline
$\omega_{\rm cdm}$   & ---   & $0.01$--$0.99$ \\
$\omega_{\rm b}$    & ---   & $0.005$--$0.1$ \\
$h$                 & ---   & $0.4$--$1.0$\\
$\tau$              & ---   & $0.01$--$0.8$ \\
$\ln(10^{10}A_{\rm s})$   & ---   & $2.7$--$4.0$ \\
$n_{\rm s}$               & ---   & $0.5$--$1.5$ \\
$m_{\rm s}$, $m_\nu$ (eV)       & 0 & $0$--$3$ \\
$N_{\rm s}$      & 0     & $0$--$7$ \\
\end{tabular}
\end{ruledtabular}
\end{table}

We use CMB data from WMAP after 7 years of
observation~\cite{Komatsu:2010fb}, as well as from the
ACBAR~\cite{Reichardt:2008ay}, BICEP~\cite{Chiang:2009xsa}, and
QuAD~\cite{Brown:2009uy} experiments.  In addition, we use the halo
power spectrum extracted from the SDSS-DR7 luminous red galaxy
sample~\cite{Reid:2009xm}.  Finally, we impose a prior on the Hubble
parameter based on the Hubble Space Telescope
observations~\cite{Riess:2009pu}.  Since recent type~Ia supernova (SN)
luminosity distance data are plagued by unresolved systematic issues
associated with the light-curve fitting methods~\cite{Kessler:2009ys},
we do not use them in our default analysis.

\begin{figure}
\includegraphics[width=7.5cm]{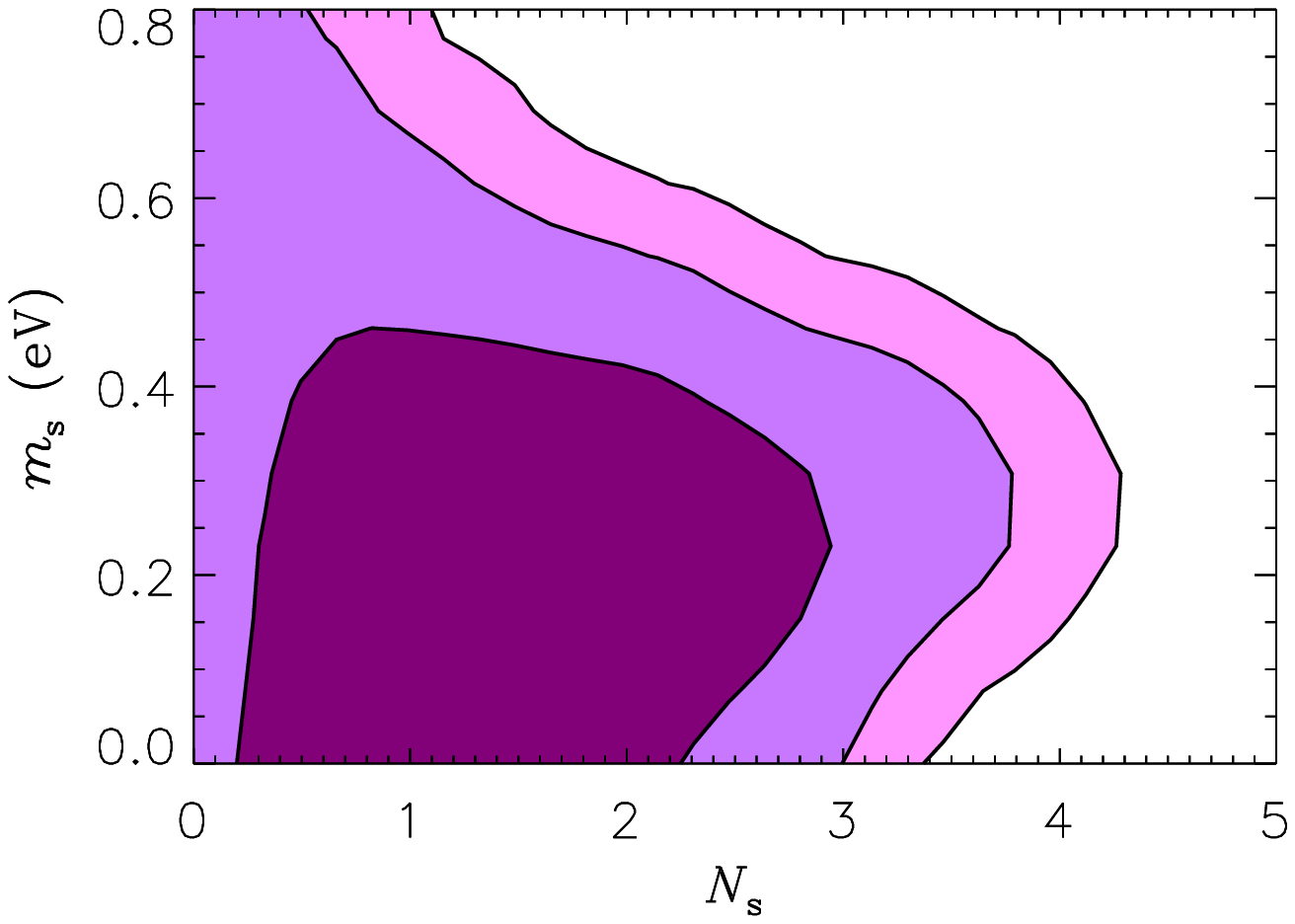}
\vskip10pt
\includegraphics[width=7.5cm]{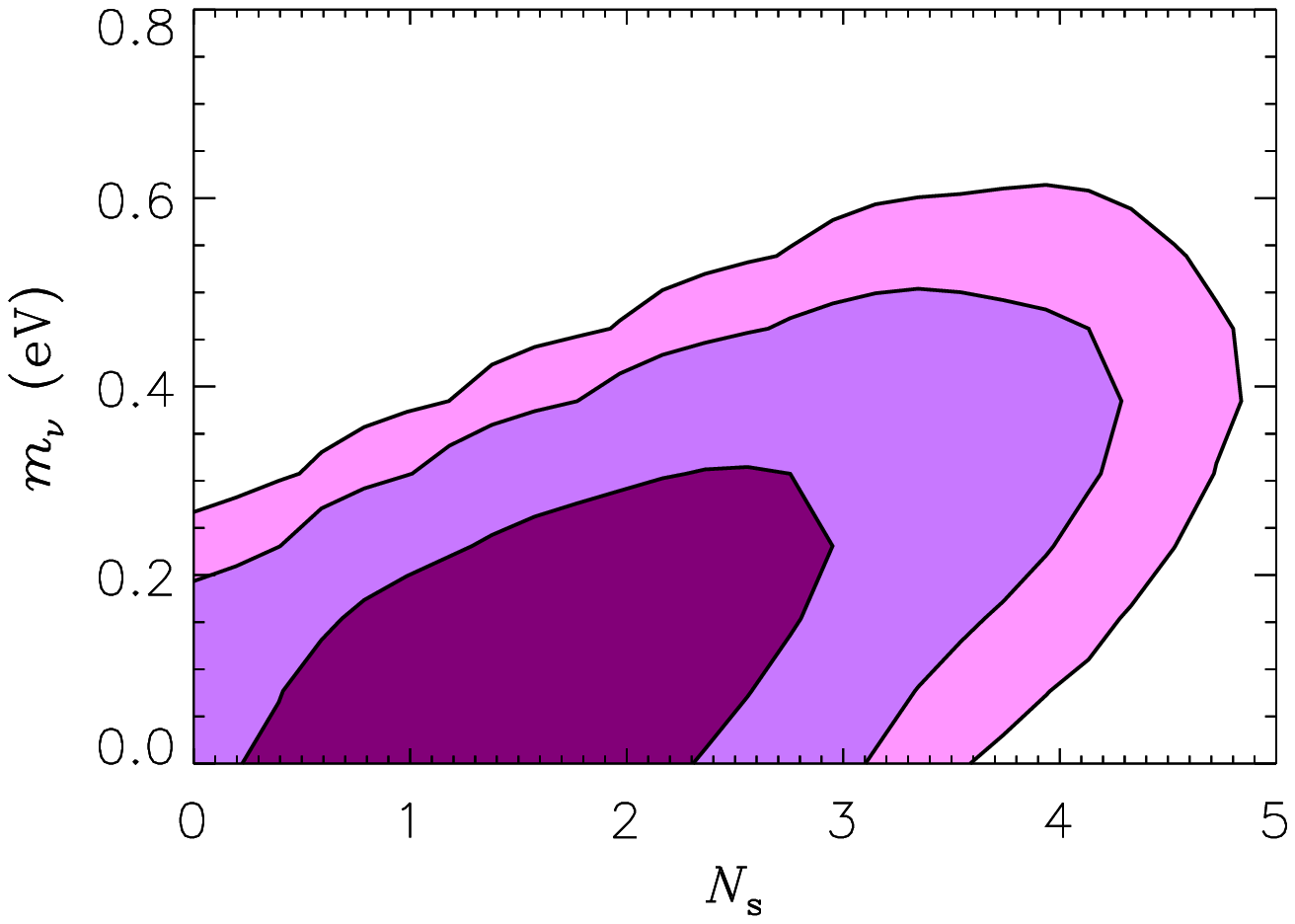}
\caption{2D marginalized 68\%, 95\% and 99\% credible regions for
  the neutrino mass and thermally excited number of degrees of freedom
  $N_{\rm s}$.  {\em Top:} The $3+N_{\rm s}$ scheme, in which ordinary
  neutrinos have $m_\nu=0$, while sterile states have a common mass
  scale $m_{\rm s}$.  {\em Bottom:} The $N_{\rm s}+3$ scheme, where
  the sterile states are taken to be massless $m_{\rm s}=0$, and 3.046
  species of ordinary neutrinos have a common mass
  $m_\nu$.\label{fig:contours}}
\end{figure}

In Figure~\ref{fig:contours}, we show 2D confidence contours for the
$3+N_{\rm s}$ and $N_{\rm s}+3$ cases. We also provide 1D
confidence intervals for $N_{\rm s}$ and $m_\nu$ or $m_s$ for
different scenarios in Table~\ref{tab:bounds}. For
$N_{\rm s}+3$ ($m_s = 0$ and
$m_\nu \neq 0$), the evidence for $N_{\rm s}>0$ is beyond 95\%, although it
is weaker in the opposite scenario $3+N_{\rm s}$ where the new
states are massive.

The evidence for $N_{\rm s} > 0$ increases if we include the
full SN sample from Ref.~\cite{Kessler:2009ys} using the {\sc
mlcs2k2}-inferred luminosity distances. However this effect is less
pronounced if adopting instead the {\sc salt-ii} light-curve fitting
method (Table~\ref{tab:bounds}). Independently of SN data, excess
radiation is preferred even by our default data~sets.

{\it BBN.}---The relatively large $N_{\rm s}$ values compatible with
this analysis are somewhat restricted by the primordial light-element
abundances.  We summarize current BBN constraints on $N_{\rm s}$ in
Table~\ref{tab:bbn}. For the $^4$He abundance we consider both the
value found by Izotov and Thuan~\cite{Izotov:2010ca}, $Y_{\rm p}^{\rm
  IT} = 0.2565 \pm 0.001 \, ({\rm stat}) \pm 0.005 \, ({\rm syst})$
and the value derived by Aver {\it et al.}~\cite{Aver:2010wq}, $Y_{\rm
  p}^{\rm A} = 0.2561 \pm 0.011$.  We adopt the deuterium
abundance of Pettini {\it et al.}~\cite{Pettini:2008mq}, $\log ({\rm
  D}/{\rm H})_{\rm p} = 4.56 \pm 0.04$ and the baryon density inferred
from our CMB+SDSS+HST analysis, $\omega_{\rm b}^{\rm CMB} = 0.02239
\pm 0.00048$. Theoretical primordial element abundances are calculated using
  \texttt{PArthENoPE}~\cite{Pisanti:2007hk}.
\begin{table}
\vspace*{-0.4cm}
\caption{1D marginalized bounds on $N_{\rm s}$ and neutrino masses.
In rows 3--6 we have used $N_s=1$ or 2 exactly. Two-tailed
    limits are minimal credible intervals.~\label{tab:bounds}}
\begin{ruledtabular}
\begin{tabular}{lllll}
Scenario& \multicolumn{2}{l}{Range for $N_{\rm s}$} &
\multicolumn{2}{l}{Range for $m_{\rm s}$ or $m_\nu$} \\
 &  68\% & 95\% &  68\% & 95\% \\
\hline
$3+N_{\rm s}$ & 0.39--2.21 & $< 3.10$ & 0.01--0.34 eV & $<0.66$~eV \\
$N_{\rm s}+3$ & 0.83--2.77 & 0.05--3.75 & $<0.22$~eV & $<0.42$~eV \\
2+3   & --- & --- & $<0.20$~eV & $<0.30$~eV \\
3+2   & --- & --- & $<0.29$~eV & $<0.45$~eV \\
1+3   & --- & --- & $<0.16$~eV & $<0.24$~eV \\
3+1   & --- & --- & $<0.35$~eV & $<0.48$~eV \\
\multicolumn{5}{l}{Including supernova data ({\sc mlcs2k2}):}\\
$3+N_{\rm s}$& 1.24--3.36 & 0.26--4.31 & 0.17--0.47~eV & 0.09--0.64~eV
\\
\multicolumn{5}{l}{Including supernova data ({\sc salt-ii}):}\\
$3+N_{\rm s}$& 0.02--1.54 & $<2.57$ & $<0.28$~eV & $<0.66$~eV \\
\end{tabular}
\end{ruledtabular}
\end{table}

It is noteworthy that independently of $^4$He, deuterium
alone also provides nontrivial limits when combined with $\omega_{\rm
  b}^{\rm CMB}$.  All of these results likely contain significant
systematic errors. We therefore refrain from a common
likelihood analysis and simply remark that both $^4$He and D somewhat
prefer $N_{\rm s}>0$, but depending on the assumed errors for $Y_{\rm
  p}$ it may be difficult to accommodate $N_{\rm s}=2$. Of course the oscillation data do
  not necessarily imply two fully
thermalized sterile neutrinos.  Indeed,
Ref.~\cite{Melchiorri:2008gq} shows that the degree of sterile state thermalization in a
 3 active + 2 sterile scenario depends strongly on the neutrino mixing and mass parameters.
Another possibility is the presence of a small lepton asymmetry, which can reduce the thermalization
efficiency~\cite{Foot:1995bm,fuller}.
Yet another option is that the oscillation data are explained
by 1 sterile state plus new
interactions~\cite{Akhmedov:2010vy}. Still, for reference we provide
mass bounds in Table~\ref{tab:bounds} for the cases of $N_{\rm s} = 1$
or 2 exactly, besides the variable $N_{\rm s}$.

\begin{table}
\caption{BBN constraints on $N_{\rm s}$, using $N_{\rm s} \geq 0$ as a prior.
  Maximum
  of the marginalised posterior and minimal 95\% credible interval (C.I.).
  \label{tab:bbn}}
\begin{ruledtabular}
\begin{tabular}{lll}
Data & Posterior max &  95\% C.I. \\
\hline
 $Y_{\rm p}^{\rm IT}$+$({\rm D}/{\rm H})_{\rm p}$  & 0.68   & $0.01$--$1.39$ \\
 $Y_{\rm p}^{{\rm A}}$+$({\rm D}/{\rm H})_{\rm p}$  & 0.69  & $< 2.42$ \\
 $({\rm D}/{\rm H})_{\rm p}$+$\omega_{\rm b}^{\rm CMB}$ & 0.49  & $< 2.12$ \\
\end{tabular}
\end{ruledtabular}
\end{table}

{\it Discussion.}---Allowing for extra radiation as a cosmological
fit parameter, current cosmological data favor additional
radiation compatible with recent hints from BBN. Assuming
ordinary neutrinos to have a common mass $m_\nu$ and the extra
radiation to be massless, evidence for $N_{\rm s}>0$ exceeds
95\%, whereas the most constraining upper limit comes from BBN. With
currently favored $^4$He and D abundances, it would be difficult to
accommodate two fully thermalized additional neutrino
states.

The usual degeneracy between extra radiation and the ordinary
neutrino mass (Figure~\ref{fig:contours}) weakens the
 neutrino mass limits, with 1D credible intervals given
in Table~\ref{tab:bounds}.

However, it is more interesting to assume essentially massless
standard neutrinos and attribute a possible mass to sterile neutrinos
($3+N_{\rm s}$ scenario). If we assume $N_{\rm s}=1$, the 95\%
allowed mass range is $m_s< 0.48$~eV. For $N_{\rm s}=2$ it is 0.45~eV
(Table~\ref{tab:bounds}), although this case would be disfavored by
BBN. For $N_{\rm s} < 1$, the 2D marginalized posterior probability
distribution has a long  tail so that $m_{\rm s}\gtrsim
1$~eV is  marginally allowed:
the fewer sterile states there are, the larger the mass they can possess.

The relatively small masses favored by cosmology are not assured to
provide good fits to the short-baseline appearance experiments---in
principle a combined analysis as in Ref.~\cite{Melchiorri:2008gq} is
desirable but complicated because of the many parameters involved.
Moreover, the degree of thermalization of the
additional states would have to be considered in detail. Two fully
thermalized states, corresponding to $N_{\rm s}=2$, are difficult to
accommodate in BBN even with the new helium abundances.

Our main message is that on present evidence, cosmology does not
exclude sterile neutrinos if they
are not too heavy and thus do not contribute excessive amounts of hot
dark matter.  Quite on the contrary, both BBN and precision
observations would happily welcome some amount of additional
radiation corresponding to around one new thermal degree of
freedom. Low-mass sterile neutrinos are one natural possibility.

Low-mass sterile neutrinos mixed with active ones can strongly modify
the neutrino signal from a core-collapse SN and
r-process nucleosynthesis in the neutrino-driven wind
\cite{Shi:1993ee, Nunokawa:1997ct, Fetter:2002xx, Keranen:2004rg,
  Cirelli:2004cz}. These effects should be studied in the
presence of two sterile flavors and CP violating phases.

The Planck spacecraft, currently taking CMB data, is expected to boost
precision determination of cosmological parameters. It is foreseen
that Planck data will constrain the cosmic radiation content at CMB
decoupling with a precision of  $\Delta N_{\rm eff} \simeq \pm0.26$ or
better, so extra radiation at the level of 2 extra species could be
detected with high significance~\cite{Perotto:2006rj,Hamann:2007sb}.
Conversely, if Planck should find $N_{\rm eff}=3\pm0.26$, this would provide strong arguments against sterile
neutrinos.  Thus, the ongoing Planck measurements will make or
break the new-found friendship between cosmology and sterile
neutrinos.

{\it Acknowledgments.}---We thank M.~Blennow,
E.~Fer\-nandez-Martinez and S. Palomares-Ruiz
for discussions. We acknowledge
use of computing resources from the Danish Center for Scientific
Computing (DCSC) and partial support by the Deutsche
Forschungsgemeinschaft, grants SFB/TR-27 and EXC~153, and by the
Italian MIUR and INFN through the ``Astroparticle Physics'' research
project.  J.H.\ is supported by a Feodor Lynen-fellowship of the
Alexander von Humboldt foundation.



\begin{thebibliography}{99}

\bibitem{Dolgov:2002wy}
  A.~D.~Dolgov,
  Phys.\ Rept.\  {\bf 370}, 333 (2002).

\bibitem{Iocco:2008va}
  F.~Iocco {\it et al.},
  Phys.\ Rept.\  {\bf 472}, 1 (2009).

\bibitem{Hamann:2007pi}
  J.~Hamann {\it et al.},
  JCAP {\bf 0708}, 021 (2007).

\bibitem{Hamann:2010pw}
  J.~Hamann {\it et al.},
  JCAP {\bf 1007}, 022 (2010).

\bibitem{GonzalezGarcia:2010un}
  M.~C.~Gonzalez-Garcia {\it et al.},
  JHEP {\bf 1008} (2010) 117.



\bibitem{Mangano:2005cc}
  G.~Mangano, {\it et al.},
  Nucl.\ Phys.\  B {\bf 729}, 221 (2005).

\bibitem{Komatsu:2010fb}
  E.~Komatsu {\it et al.},
  arXiv:1001.4538.

\bibitem{Reid:2009nq}
  B.~A.~Reid {\it et al.},
  JCAP {\bf 1001}, 003 (2010).

\bibitem{Dunkley:2010ge}
  J.~Dunkley {\it et al.} (Atacama Cosmology Telescope
  Collaboration),
  arXiv:1009.0866.

\bibitem{Simha:2008zj}
  V.~Simha and G.~Steigman,
  JCAP {\bf 0806}, 016 (2008).

\bibitem{Izotov:2010ca}
  Y.~I.~Izotov and T.~X.~Thuan,
  Astrophys.\ J.\  {\bf 710}, L67 (2010).

\bibitem{Aver:2010wq}
  E.~Aver {\it et al.},
  JCAP {\bf 1005}, 003 (2010).

\bibitem{Chun:2000jr}
  E.~J.~Chun {\it et al.},
  Phys.\ Rev.\  D {\bf 62}, 095013 (2000).

\bibitem{Raffelt:1999tx}
  G.~G.~Raffelt,
  Ann.\ Rev.\ Nucl.\ Part.\ Sci.\  {\bf 49}, 163 (1999).

\bibitem{Hannestad:2005df}
  S.~Hannestad {\it et al.},
  JCAP {\bf 0507}, 002 (2005).

\bibitem{Pastor:2008ti}
  S.~Pastor {\it et al.},
  Phys.\ Rev.\ Lett.\  {\bf 102}, 241302 (2009).

\bibitem{Kainulainen:1990ds}
  K.~Kainulainen,
  Phys.\ Lett.\  B {\bf 244}, 191 (1990).

\bibitem{Aguilar:2001ty}
  A.~Aguilar {\it et al.} (LSND Collaboration),
  Phys.\ Rev.\  D {\bf 64}, 112007 (2001).

\bibitem{GonzalezGarcia:2007ib}
  M.~C.~Gonzalez-Garcia and M.~Maltoni,
  Phys.\ Rept.\  {\bf 460}, 1 (2008).

\bibitem{Strumia:2002fw}
  A.~Strumia,
  Phys.\ Lett.\ B {\bf 539}, 91 (2002).

\bibitem{Cirelli:2004cz}
  M.~Cirelli {\it et al.},
  Nucl.\ Phys.\  B {\bf 708}, 215 (2005).

\bibitem{AguilarArevalo:2007it}
  A.~A.~Aguilar-Arevalo {\it et al.} (MiniBooNE Collaboration),
  Phys.\ Rev.\ Lett.\  {\bf 98}, 231801 (2007).

\bibitem{AguilarArevalo:2008rc}
  A.~A.~Aguilar-Arevalo {\it et al.} (MiniBooNE Collaboration),
  Phys.\ Rev.\ Lett.\  {\bf 102}, 101802 (2009).

\bibitem{AguilarArevalo:2009xn}
  A.~A.~Aguilar-Arevalo {\it et al.} (MiniBooNE Collaboration),
  Phys.\ Rev.\ Lett.\  {\bf 103}, 111801 (2009).

\bibitem{MB2010}
  R.~Van de Water for the MiniBooNE Collaboration,
  ``Updated anti-neutrino oscillation results from MiniBooNE,''
  (Neutrino 2010,
  14-19 June 2010, Athens, Greece), see
  http://indico.cern.ch/event/73981.

\bibitem{Karagiorgi:2009nb}
  G.~Karagiorgi {\it et al.},
  Phys.\ Rev.\  D {\bf 80}, 073001 (2009);
  Erratum ibid.\  D {\bf 81}, 039902 (2010).

\bibitem{Sterile2010}
  G.~Karagiorgi,
  ``Toward Solution of the MiniBooNE and LSND Anomalies,''
  (Neutrino 2010, see Ref.~\cite{MB2010}).

\bibitem{Akhmedov:2010vy}
  E.~Akhmedov and T.~Schwetz,
  arXiv:1007.4171.

\bibitem{Giunti:2010wz}
  C.~Giunti and M.~Laveder,
  arXiv:1005.4599.

\bibitem{Dodelson:2005tp}
  S.~Dodelson {\it et al.},
  Phys.\ Rev.\ Lett.\  {\bf 97}, 04301 (2006).

\bibitem{Melchiorri:2008gq}
  A.~Melchiorri {\it et al.},
  JCAP {\bf 0901}, 036 (2009).

\bibitem{Acero:2008rh}
  M.~A.~Acero and J.~Lesgourgues,
  Phys.\ Rev.\  D {\bf 79}, 045026 (2009).

\bibitem{DiBari:2001ua}
  P.~Di Bari,
  Phys.\ Rev.\  D {\bf 65}, 043509 (2002) and
  Addendum ibid.\  D {\bf 67}, 127301 (2003).

\bibitem{Foot:1995bm}
  R.~Foot, R.~R.~Volkas,
  Phys.\ Rev.\ Lett.\  {\bf 75 }, 4350  (1995).

\bibitem{fuller}
  K.~Abazajian {\it et al.},
  Phys.\ Rev.\  D {\bf 64}, 023501 (2001).

\bibitem{Lewis:2002ah}
  A.~Lewis and S.~Bridle,
  Phys.\ Rev.\  D {\bf 66}, 103511 (2002).

\bibitem{Reichardt:2008ay}
  C.~L.~Reichardt {\it et al.},
  Astrophys.\ J.\  {\bf 694}, 1200 (2009).

\bibitem{Chiang:2009xsa}
  H.~C.~Chiang {\it et al.},
  Astrophys.\ J.\  {\bf 711}, 1123 (2010).

\bibitem{Brown:2009uy}
  M.~L.~Brown {\it et al.} (QUaD collaboration),
  Astrophys.\ J.\  {\bf 705}, 978 (2009).

\bibitem{Reid:2009xm}
  B.~A.~Reid {\it et al.},
  Mon.\ Not.\ Roy.\ Astron.\ Soc.\  {\bf 404}, 60  (2010).

\bibitem{Riess:2009pu}
  A.~G.~Riess {\it et al.},
  Astrophys.\ J.\  {\bf 699}, 539 (2009).

\bibitem{Kessler:2009ys}
  R.~Kessler {\it et al.},
  Astrophys.\ J.\ Suppl.\  {\bf 185}, 32 (2009).

\bibitem{Pettini:2008mq}
  M.~Pettini {\it et al.},
  Mon.\ Not.\ Roy.\ Astron.\ Soc.\  {\bf 391}, 1499 (2008).

\bibitem{Pisanti:2007hk}
  O.~Pisanti {\it et al.},
  Comput.\ Phys.\ Commun.\  {\bf 178} (2008) 956.

\bibitem{Shi:1993ee}
  X.~Shi and G.~Sigl,
  Phys.\ Lett.\  B {\bf 323}, 360 (1994);
  Erratum ibid.\  B {\bf 324}, 516 (1994).

\bibitem{Nunokawa:1997ct}
  H.~Nunokawa {\it et al.},
  Phys.\ Rev.\  D {\bf 56}, 1704 (1997).

\bibitem{Fetter:2002xx}
 J.~Fetter {\it et al.},
 Astropart.\ Phys.\  {\bf 18}, 433 (2003).

\bibitem{Keranen:2004rg}
  P.~Ker\"anen {\it et al.},
  Phys.\ Lett.\  B {\bf 597}, 374 (2004).

\bibitem{Perotto:2006rj}
  L.~Perotto {\it et al.},
  JCAP {\bf 0610}, 013 (2006).

\bibitem{Hamann:2007sb}
  J.~Hamann {\it et al.},
  JCAP {\bf 0803}, 004 (2008).

\end{thebibliography}
\end{document}